# A Mediation Framework for Mobile Web Service Provisioning

Satish Narayana Srirama[1], Matthias Jarke[1,2], Wolfgang Prinz[1,2]
[1]*RWTH Aachen, Informatik V*
*Ahornstr.55, 52056 Aachen, Germany*
[2]*Fraunhofer FIT*
*Schloss Birlinghoven, 53754 Sankt Augustin, Germany*
*{srirama, jarke}@cs.rwth-aachen.de*
*wolfgang.prinz@fit.fraunhofer.de*

## Abstract

*Web Services and mobile data services are the newest trends in information systems engineering in wired and wireless domains, respectively. Web Services have a broad range of service distributions while mobile phones have large and expanding user base. To address the confluence of Web Services and pervasive mobile devices and communication environments, a basic mobile Web Service provider was developed for smart phones. The performance of this Mobile Host was also analyzed in detail. Further analysis of the Mobile Host to provide proper QoS and to check Mobile Host's feasibility in the P2P networks, identified the necessity of a mediation framework. The paper describes the research conducted with the Mobile Host, identifies the tasks of the mediation framework and then discusses the feasible realization details of such a mobile Web Services mediation framework.*

## 1. Introduction

With the current generation of mobile devices like smart phones, PDAs and other consumer devices, the wireless market is expanding very fast. People are using such high-end mobile phones and devices for wide range of applications like mobile banking, location based services, e-learning etc. The situation also brings out a large scope and demand for software applications for such high-end mobile devices.

Service Oriented Architecture is the latest trend in information systems engineering. It is a component model, presenting an approach for building distributed systems. SOA delivers application functionality as services to end-user applications and other services, bringing the benefits of loose coupling and encapsulation to the enterprise application integration. SOA is not a new notion and many technologies like CORBA and DCOM are at least partly represent this idea. Web Services are newest of these developments and by far the best means of achieving SOA.

Web Services and its protocol stack are based on open standards and are widely accepted over the internet community. Web Services have wide range of applications and range from simple stock quotes to pervasive applications using context awareness like weather forecasts, map services etc. The biggest advantage of Web Services lies in its simplicity in expression, communication and servicing. The componentized architecture of Web Services also makes them reusable, thereby reducing the development time and costs. The fine-grained atomic services can also be orchestrated to coarse-grained business services, simplifying the coupling between business processes. [1]

Web Services have a broad range of service distributions and on the other hand cellular phones have large user base and it is increasing day by day. Most recently, with the achieved high data transmission rates in cellular domain, with interim and third generation mobile communication technologies like GPRS, EDGE and UMTS [2], mobiles are also being used as Web Service clients and providers, bridging the gap between the wireless networks and the stationery IP data networks. Combining Web Services with mobile technology brings us a new trend and lead to manifold opportunities to mobile operators, wireless equipment vendors, third-party application developers, and end users. [3, 4]

During one of our previous projects, a small mobile Web Service provider ("Mobile Host") has been developed for resource constrained devices like smart phones. The detailed performance analysis conducted with the Mobile Host showed that the processing capability, time frames are very much within the acceptable levels. But the QoS analysis of the Mobile Host with main interest on security aspects showed that not all the standard WS* specifications can be applied to the Mobile Host. A middleware framework or proxy is required for the proper QoS provisioning. We also have extended the study to check the feasibility of this Mobile Host in the P2P networks. The analysis suggested that the mediation framework also need to provide some of the P2P features for the Mobile Host.

The paper first explains the Mobile Host and the research being conducted in the QoS and P2P domains in section 2. Section 3 identifies the deployment scenario, tasks and realization details of the mobile Web Services mediation framework. Section 4 concludes the paper with future research directions.

## 2. Mobile Web Service provisioning

Mobile Host is a light weight Web Service provider built for resource constrained devices like smart phones. It has been developed as a Web Service handler built on top of a normal Web server. The Web Service requests sent by HTTP tunneling are diverted and handled by the Web Service handler. The Mobile Host was developed in PersonalJava on a SonyEricsson P800 smart phone. The memory imprint of our fully functional prototype is only ~130 KB. Open source kSOAP2 was used for creating and handling the SOAP messages [4]. An application scenario of the Mobile Host where the pictures taken by the mobile and location details are provided as Web Service to the potential clients is shown in figure 1.

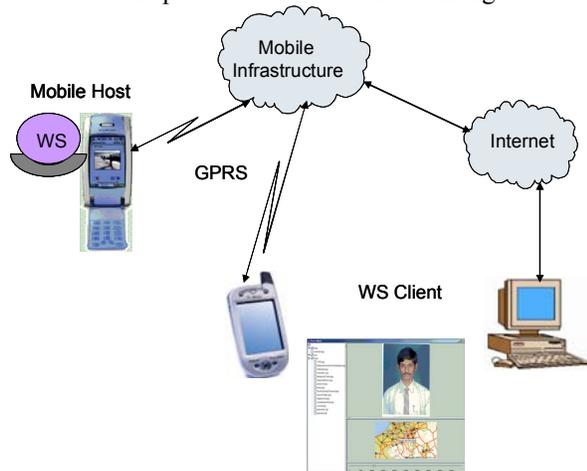

**Figure 1. Mobile photo album scenario**

The detailed performance evaluation of this Mobile Host clearly showed that service delivery as well as service administration can be done with reasonable ergonomic quality by normal mobile phone users. As the most important result, it turns out that the WS processing time at the Mobile Host is only a small fraction of the total Web Service invocation cycle time (<10%) and rest all being transmission delay. This makes the performance of the Mobile Host directly proportional to achievable data transmission rates. So the new developments and higher data transmission rates achieved in mobile communication technologies with 3G, 4G make the Mobile Host soon realizable in commercial environments.

After the initial feasibility analysis, the QoS aspects of Mobile Host are studied with main concern at the security implications. The WS-Security [5] specification from OASIS is the core element in Web Service security realm in wired networks. It provides ways to add security headers to SOAP envelopes, attach security tokens and credentials to a message, insert a timestamp, sign the messages, and encrypt the message. The detailed analysis of adapting the WS-Security to the Mobile Host suggested that not all of this specification can be applied to the Mobile Host. The specification was beyond the resource capabilities of today's smart phones. With the security study we are recommending that the best way of securing SOAP messages in mobile Web Service provisioning is to use AES symmetric encryption with 256 bit key for encrypting the message and RSAwithSHA1 to sign the message. The symmetric keys are to be exchanged using RSA 1024 bit asymmetric key exchange mechanism. The cipher data and the keys are to be incorporated into the SOAP message according to WS-Security specification. [7]

But in general any potential client can follow full WS-Security standard and hence to preserve the interoperability of Web Services also at Mobile Host, the messages are to be transformed at some intermediary node. The transformation of the message may also be required to achieve proper scalability for the Mobile Host. This pushes the necessity for some mediation framework as the legitimate intermediary in the mobile Web Service invocation cycle. Our current study in this domain concentrates at realizing an Enterprise Service Bus (ESB) [9] based "Mobile Web Services Mediation Framework" (MWSMF), which maintains the individual user profiles, personalization settings and context sensitive information.

Apart from this, the feasibility of the Mobile Host in P2P networks is also being studied. The analysis suggests that the P2P is not just increasing the application scope of the Mobile Host but also provides many technical advantages to the Mobile Host. Every Mobile Host in the internet or operator proprietary network needs some means of addressing and accessing mechanism. Generally they can be identified with public IP, a feature provided by very few mobile operators today. This is the major hindrance for the commercial viability of the Mobile Host. With the P2P network in place the need for the Public IP can be eliminated and the mobiles can be addressed with unique peer ID. Each device in the P2P network is associated with the same peer ID, even though the peers can communicate with each other using the best of the many network interfaces supported by the devices like Ethernet, WiFi etc. [7]

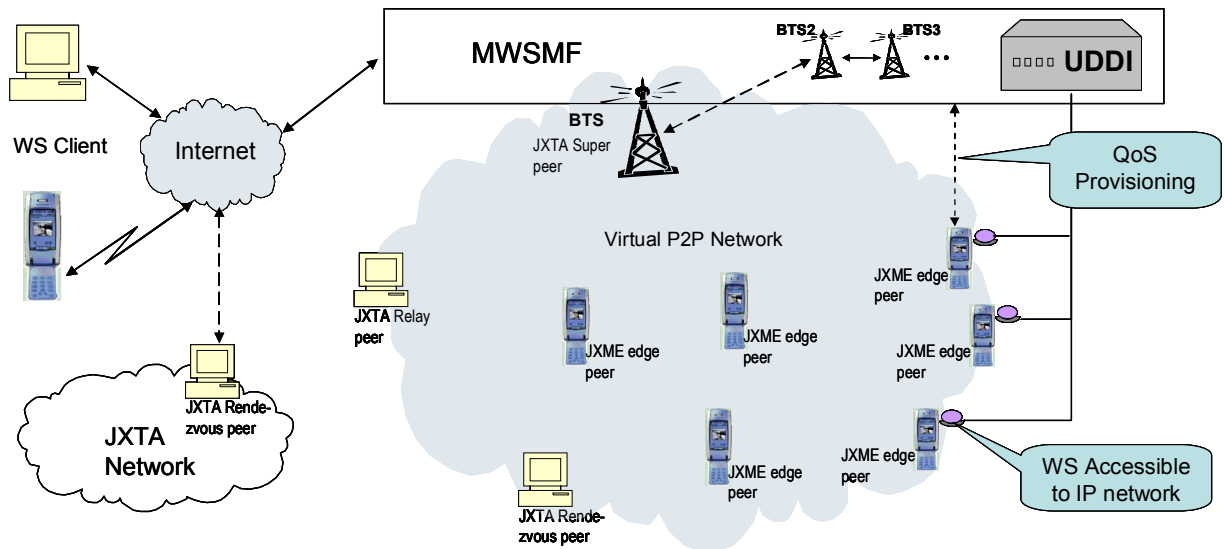

**Figure 2. The Mobile Web Services Mediation Framework deployment scenario**

Moreover, the P2P also offers better means of service discovery for huge number of Web Services possible with Mobile Hosts. The Web Services deployed on the Mobile Host can be advertised in the P2P network as the JXTA modules [6]. The module advertisements are searchable and thus the Web Services can be found as standard JXTA services. The module advertisements together represent a combination of UDDI in a sense of publishing and finding service description and WSDL in a sense of defining transport binding to the service [10, 11]. The means of accessing Web Service deployed in the JXTA network is also being studied. The SOAP messages are transmitted across JXTA pipes. The message transformation service, JXTA discovery service of the MWSMF comes in handy at this process.

For establishing the mobile P2P network the Mobile Hosts and the mobile Web Service clients must connect to some super peer. Super peer helps in maintaining the routing information and searching the resources and services in a P2P network. The mediation framework should posses the super peer functionality, and should lead the JXTA network to the mobile operator proprietary network.

## 3. Mobile Web Services Mediation Framework

The proposed deployment scenario of the MWSMF is shown in figure 2. The mediation framework is established as an intermediary between the WS clients and the Mobile Hosts in JXTA network. The virtual P2P network is established in the mobile operator network with one of the node in operator proprietary network, acting as a JXTA super peer. The super peer can exist at Base Transceiver Station (BTS) and can be connected to other base stations extending the JXTA network into the mobile operator network. The mobile terminals use JXME, a light version of JXTA for mobile devices. The WS clients can access the deployed services across MWSMF and JXTA network.

External WS clients can also directly access the Web Services deployed on the Mobile Hosts, as long as the Mobile Hosts are provided with public IPs and the Web Services are published with the UDDI registry at the mediation framework. Thus the MWSMF also acts as external gateway from Internet to the mobile P2P network.

### 3.1. Features of the mediation framework

Based on the study and analysis, addressed in the previous section, the tasks of the mediation framework are identified. The mediation framework should support the transformation and routing of the Web Service messages. The routing should be based on content. It should help in providing QoS for the Mobile Hosts and mobile Web Service clients. From security front, MWSMF should provide end point security by providing identity, which helps in achieving proper authentication and authorization. From the scalability front it should support the transformation of messages between the complete WS* specifications and the specifications feasible for mobile Web Service provisioning.

MWSMF should also support automatic startup of the Mobile Hosts. Generally hand-held devices have many resource limitations like low computation capacities, limited storage capacities, limited battery power etc. So to conserve these resources, the Mobile Host features of the smart phones can turn-on only at the client request. The MWSMF can identify when the request is for particular mobile phone, using profiles, and can send a SMS message based on specific protocol, which activates the Mobile Host.

Moreover, with the introduction of Mobile Host in P2P, the MWSMF should provide the functionality of super peer for the JXME edge peers. The Mediation framework should also support advanced and context aware search of Modules along with support for composition, orchestration and choreography of services.

Even though the Web Services deployed on the Mobile Host are advertised in JXTA network and identifiable with peer ID, the smart phones can still posses the public IP feature. The Web Services can thus be published with the UDDI registry at the MWSMF. Any external client can search the registry and can access the Web Services directly. So the mediation framework should support the access of Web Services both across P2P and standard WS protocols. Thus the mediation framework acts as an external gateway to the P2P network.

### 3.2. Realization of the MWSMF

ESB [9] is the emerging infrastructure component for realizing SOA. ESBs provide the dedicated infrastructure with the capability to route and transport the service requests to the respective/correct service providers. The infrastructure is established based on basic and emerging WS* specifications and standards and also exploit message oriented middleware, intelligent routing, and transformation. Many products have already hit the market like Sonic Software, Artix, Cape Clear etc. [9] Most of these products are based on proprietary message middleware or extensions to EAI architectures. SUN has defined JSR 208, Java business integration specification, and ESB products like Service Mix are based on this specification.

To realize the MWSMF, we are trying to adapt the ServiceMix to the identified features and requirements. Proper adapters will be defined for Mobile Host and P2P support. A standard based UDDI registry will also be mapped to the mediation framework.

## 4. Conclusion

This paper addressed the necessity of middleware for mobile Web Service provisioning from smart phones. The features, realization details and deployment scenario of such a mobile Web Services mediation framework are discussed. The approach throws lot of scope for further research.

First steps on the technical side include the realization of the mobile P2P network and MWSMF, with little/reasonable performance penalty on the Mobile Host. Later steps include optimizing the complete scenario with features like advanced and context aware search of services and increasing the scalability of this approach.

## 5. Acknowledgement

The work is supported by German Research Foundation (DFG) as part of the Graduate School "Software for Mobile Communication Systems" at RWTH Aachen University. The authors would also like to thank R. Levenshteyn and M. Gerdes of Ericsson Research for their help and support.